\documentstyle[preprint,prc,aps]{revtex}

\date{\today}
\begin{document}
\title{Scalar meson exchange and the baryon spectra\thanks{to appear in
the Proceedings of the Third International Conference on
{\it {Quark Confinement and the Hadron Spectrum III}}, Jefferson Lab.
(Newport News, USA) June 7-12, 1998 , World Scientific (Singapore) }}
\author{P. Stassart and Fl. Stancu}
\address{Universit\'{e} de Li\`ege, Institut de Physique B.5, Sart Tilman,
B-4000 Li\`ege 1, Belgium}
\author{J.-M. Richard}
\address{Institut des Sciences Nucl\'eaires, Universit\'e Joseph
Fourier\\IN2P3-CNRS, 53, Avenue des Martyrs\\F-38026 Grenoble Cedex, France}

\maketitle
\everymath={\displaystyle}

\vspace{1cm}

\begin{abstract}
We explore the role of a scalar meson exchange interaction between quarks in a
semirelativistic constituent quark model where the quarks are subject to a
linear confinement. We search for a variational solution and show that the gap
between the $N=1$ band and the ground state $N=0$ band increases with the
strength of the scalar meson exchange interaction potential. This result has
good implications on the description of the low-lying baryon masses, especially
on the Roper vs. negative-parity resonances ordering.\par
\end{abstract}

\vspace{1cm}
It has been suggested \cite{MA84,GL96a} that beyond the scale of spontaneous
chiral sypmmetry breaking, nonstrange and strange baryons can be viewed as
systems of three quarks interacting via exchange of Goldstone bosons
(pseudoscalar mesons). It has also been shown that a properly parametrized
interaction provides good baryon spectra with a correct order of positive and
negative parity levels both in a nonrelativistic \cite{GL96b} or a
semirelativistic \cite{GL97} treatment. In both parametrizations the
pseudoscalar meson exchange interaction has two distinct parts: a long-range
Yukawa potential tail and a short-range part having opposite sign as
compared to
the Yukawa potential tail. It is the latter which plays a major role in
describing the baryon spectra in the frame of Goldstone boson exchange (GBE)
models.\par
Although the models \cite{GL96a,GL96b,GL97} are thought to be a consequence of
the spontaneous chiral symmetry breaking, the chiral partner of the pion, the
sigma meson, is not considered explicitly. One can think of having mocked
up its
contribution in the parameters of the Hamiltonian \cite{GL96b,GL97}, e.g.
in the
regularization parameter of the short-range spin-spin term. The price which
could have been paid is the large role played by the $\eta'$ meson 
exchange which comes
into the interaction with a strength $g_0^2/4\pi$ about equal \cite{GL97} or
larger \cite{GL96b} than the strength $g_8^2/4\pi$ associated with the
pseudoscalar octet $\left(\pi, K, \eta\right)$.\par
Here we study the explicit role of a sigma-exchange interaction by considering
the following Hamiltonian
\begin{equation}
H_0 = \sum\limits_{i}^{} \left(m_i^2 + p_i^2\right)^{1/2} + \frac{1}{2}
\sqrt{\sigma} \sum\limits_{i<j}^{} |\vec{r}_i - \vec{r}_j| -
\frac{g_{\sigma}^2}{4\pi}
\sum\limits_{i<j}^{} \frac{exp \left[- \mu_{\sigma} |\vec{r}_i -
\vec{r}_j|\right]}{|\vec{r}_i - \vec{r}_j|}
\end{equation}
where the second term is the confinement potential with a strength tension
\cite{CA83}
\begin{equation}
\sqrt{\sigma} = {1 \ GeV \ fm^{-1}}
\end{equation}
$\mu_{\sigma}$ = 600 MeV is the sigma meson mass and $g_{\sigma}^2/4\pi$ is the
coupling constant sigma-quark, taken as a variable parameter.
From the pion-nucleon coupling constant $g_{\pi
NN}^2/4\pi \simeq$ 14, one obtains a pion-quark coupling constant $g_{\pi
qq}^2/4\pi$ = 0.67 which has been used in \cite{GL96b,GL97}. Assuming a
sigma-nucleon coupling $g_{\sigma NN}^2/4\pi \simeq$ 8 \cite{MA87} one obtains, 
by scaling, 
 $g_{\sigma qq}^2/4\pi \simeq$ 0.4. In the following we study the
role of the $\sigma$-meson exchange by considering values of $g_{\sigma
qq}^2/4\pi$ up to 0.2. A qualitative argument that the $\sigma$-meson
contributes attractively in (1) is that we view it as a two correlated
pions, as
in the nucleon-nucleon interaction. Moreover, one can consider more
sophisticated arguments related to the differences appearing in the scalar and
pseudoscalar propagators \cite{JA89} in a Nambu-Jona-Lasinio model. From there
one can argue that the Fourier transform would lead to a Yukawa + a contact
term for a scalar and a Yukawa - a contact term, for a pseudoscalar exchange,
as in Refs. \cite{GL96b,GL97}. Due to the addition of the long and short range
contributions in the scalar case,
here we consider a simplified form of the $\sigma$-exchange
potential, containing a single term as in (1).\par
The variational wave function has the form
\begin{equation}
\psi_{n}\left(\vec{r}_{12},\vec{r}_{13},\vec{r}_{23}\right) = F_{123}
\left[\prod\limits_{i<j}^{}
f\left(r_{ij}\right)\phi_n\left(\vec{r}_{12},\vec{r}_{13},
\vec{r}_{13}\right)\right].
\end{equation}
where $f$ and $F_{123}$ are two- and three-body parts of the 
ground state wave function and $\Phi_{n}$ with $n\neq 0$ contain
orbital excitations \cite{SS85}. The function $f$ is parametrized as
\begin{equation}
\begin{array}{c}
f\left(r\right) = r^{\delta} exp\left\{-W\left(r\right)\gamma_1r - \left[1 -
W\left(r\right)\right]\gamma_{1.5}r^{1.5}\right\},\\

\vspace{3mm}

W\left(r\right) = \frac{1 + exp\left(-r_0/a\right)}{1 + exp\left[\left(r -
r_0\right)/a\right]}
\end{array}
\end{equation}
The three-body part is chosen as
\begin{equation}
F_{123} = \left[1 - \beta \sqrt{\sigma} \left(\sum\limits_{i}^{} r_{i4} -
\frac{1}{2} \sum\limits_{i < j}^{} r_{ij}\right)\right].
\end{equation}

The quantities 
$\gamma_1$, $\gamma_{1.5}$, $a$, $r_0$, $\delta$ and $\beta$ are  
variational parameters. In the minimization procedure we found
that the ground state expectation value of (1) varies smoothly
with all these parameters but $\delta$. It is quite natural
because $\delta$ cares for the behaviour of the wave function
near the origin, typical for the solution of a 
relativistic equation with a singular potential.
The larger the coupling constant $g_{\sigma}^2/4\pi$ the 
more difficult in reaching the minimum. The expectation
value for the first
two excited states are obtained
with the optimal values of the ground state parameters.
The result is exhibited in the table below.\par

\vspace{1cm}

\begin{center}
\begin{tabular}{|l|l|l|l|}
\hline
$g_{\sigma}^2/4\pi$\,\, & ground st. & neg.parity & radial excit. \\
\hline
\, 0.00 &  \,\,\, 940   &   \,\,\, 1287  &   \,\,\, 1446 \\ 
\, 0.05 &  \,\,\, 940   &   \,\,\, 1309  &   \,\,\, 1444 \\
\, 0.10 &  \,\,\, 940   &   \,\,\, 1333  &   \,\,\, 1444 \\
\, 0.15 &  \,\,\, 940   &   \,\,\, 1380  &   \,\,\, 1445 \\
\, 0.20 &  \,\,\, 940   &   \,\,\, 1431  &   \,\,\, 1446 \\
\hline
\end{tabular}
\end{center}

\vspace{1cm}

One can see that the mass difference between the first radially
excited state and the negative parity state is positive
up to the coupling constant value of 0.20 but tends to
vanish. Note also that the difference 
between the radially excited state and the ground state
remains practically constant as a function of the coupling
constant, while the mass difference between the orbitally excited
state and the ground state increases with $g_{\sigma}^2/4\pi$.
The reason for this is that at $g_{\sigma} \neq 0 $ all $s$
states are lowered with respect to the $p$ states.
The wave function of the latter is small around the origin,
so that it reduces some of the attraction in the expectation value.\par  
Preliminary results indicate that the mass difference between the first 
radially excited state and the negative parity state becomes  
negative for  $g_{\sigma}^2/4\pi > 0.20$.
This is precisely the desired behaviour
for reproducing the correct order of the experimental spectrum, 
as in Ref. \cite{GL96b}. These calculations indicate that a model
incorporating a potential, whose Laplacian is negative in a  
region around the origin, can yield the right ordering
of the lowest positive and negative parity states. 
For the potential of Eq. (1) such a situation is achieved
for suitable values of the coupling constant $g_{\sigma}^2/4\pi$.
The Laplacian is related to the concavity of the two-body potential 
as discussed in Ref. \cite{JMR}. \par 
The conclusion is that
Goldstone boson exchange models 
should include explicitly the chiral partner of
the pseudoscalar mesons. It might be misleading to mock up
its effect in some nontrivial parameters of the model.
We are presently studying the case of a potential with both
scalar and pseudoscalar terms. However the question still
remains about the role of the chromomagnetic interaction.\par 
Finally, the scalar exchange term, which we have shown to be
important for the level ordering, might also genearate a new
spin-orbit contribution. It would be interesting to see whether or not
that contribution could significantly improve our present
knowledge of the spin-orbit problem in baryons \cite{GRO}.

\end{document}